\documentclass[a4paper]{jpconf}
\usepackage{graphicx}
\begin{document}
\title{Electron capture rates in stars studied with heavy ion charge exchange reactions}

\author{C.A. Bertulani}

\address{Department of Physics and Astronomy, Texas A\&M University-Commerce, Commerce, TX 75429, USA }

\ead{carlos.bertulani@tamuc.edu}

\begin{abstract}
Indirect methods using nucleus-nucleus reactions at high energies (here, high energies mean $\sim$ 50 MeV/nucleon and higher) are now routinely used to extract information of interest for nuclear astrophysics. This is of extreme relevance  as many of the nuclei involved in stellar evolution are short-lived. Therefore,  indirect methods became the focus of recent studies carried out in major  nuclear physics facilities. Among such methods, heavy ion charge exchange is thought to be a useful tool to infer Gamow-Teller matrix elements needed to describe electron capture rates in stars and also double beta-decay experiments.  In this short review, I provide a theoretical guidance based on a simple reaction model for charge exchange reactions.
\end{abstract}

\section{Introduction}
Supernovae explosions may provide a scenario for the production of heavy elements in the Universe, although it is uncertain what are the mechanisms driving such explosions \cite{Heg03}. The uncertainties arise from several possibilities, including how fast nuclear reactions occur in their environments. E.g., electron-capture on neutron-rich nuclei beyond mass number  60  is thought to become important as the density increases during core-collapse supernovae \cite{Hix03}.  One cannot access these reaction rates directly in laboratory experiments. Additional problems are, e.g.,  the stellar enhancement due to thermal population of excited states.  Moreover, in the medium nuclear mass range, accurate shell-model calculations are difficult and theoretical methods employing mean-field techniques have been introduced which include large uncertainties.
To check their validity, theoretical calculations must be tested against experiment. Another even more difficult problem arises because laboratory-based experiments do not reproduce the conditions (density and temperature) present in stellar environments \cite{Hix03,Pin00}. Thus the numerous electron capture reactions occurring in stars need coordinated efforts involving theory and experiments.

To fill in parts of our knowledge gap, (p,n), ($^3$He,t) and heavy ion charge-exchange reactions have been used to extract Gamow-Teller matrix elements inherent to electron capture rates.  Experimental analysis often use the formalism popularized in Refs.  \cite{Goo80,Pet80,Tad87,Ost92}, which relates the  forward scattering charge-exchange formula with the reduced matrix elements for non spin-flip Fermi, $B(F)$, and spin-flip Gamow-Teller, $B(GT)$, matrix elements, i.e., 
\begin{equation}
{d\sigma\over d\Omega}(\theta=0^\circ)=\left( \mu \over 2\pi \hbar\right)^2 {k_f \over k_i} N_D|J_{\sigma\tau}|^2 \left[ B(GT) + {\cal C}B(F) \right], \label{tadeucci}
\end{equation}
where  $N_D$ is a distortion factor
(accounting for initial and final state interactions),
$J_{\sigma\tau}$ is the Fourier transform of the GT part of the effective
nucleon-nucleon interaction, ${\cal C} = \left| J_\tau/J_{\sigma\tau}\right|^2$, and  $B(F)= A_J | \langle f ||\sum_k  \tau_k^{(\pm)} || i \rangle |^2$, and 
$B(GT)= A_J | \langle f ||\sum_k \sigma_k \tau_k^{(\pm)} || i \rangle |^2$ with $A_J=(2J_f+1)/(2J_i+1)$. Eq. \ref{tadeucci} is purely empirical and there are strong evidences that it works reasonably well for most reactions of interest. It has been used indiscriminately in the analysis of charge exchange reactions, although it has been also shown that it fails in few cases. It lacks a solid theoretical basis and should be used with caution to reach the accuracy needed for the electron capture response functions \cite{Ber93,BL97}.  In view of the great relevance of the ongoing experimental campaign in radioactive beam facilities, I will show how Eq. \ref{tadeucci} arises from a very simple model and what are its limitations.

\section{Charge-exchange processes with heavy ion reactions}
As shown in Fig.  \ref{fig1} left panel, charge-exchange reactions can proceed via virtual pion and rho exchange in one or two step processes (a, b), whereas a physical nucleon exchange always requires at least a two step process (c). Therefore, pion and rho exchange is a more efficient way to induce charge exchange reactions with heavy ions, unless the reaction occurs at low energies of about 50 MeV/nucleon and lower, as demonstrated by Lenske et al. \cite{Lens89}. This fact also allows the reaction mechanism to be much simpler to handle because nucleon exchange is much harder to describe theoretically than pion or rho exchange.
\subsection{General formalism}
The DWBA scattering amplitude for inelastic processes in nucleus-nucleus
collisions is given by \cite{BD04}%
\begin{equation}
f_{aA\rightarrow bB}\left(  \theta\right)  =-\frac{m_{0}}{2\pi\hbar^{2}%
}\left\langle \chi_{\mathbf{k}^{\prime}}^{(-)}(\mathbf{R}\,)\phi
_{f}\,(\mathbf{r}\,)|V(\mathbf{R},\mathbf{r}\,)|\chi_{\mathbf{k}}%
^{(+)}(\mathbf{R})\phi_{i}(\mathbf{r}\,)\right\rangle \label{tfi1}%
\end{equation}
where $\chi_{\mathbf{k}^{\prime}\mathbf{,k}}^{(\pm)}\,$\ are the
(outgoing/incoming) distorted scattered waves for the c.m. motion of the two
nuclei with reduced mass $m_{0}$ and $\,\phi_{i,f}(\mathbf{r\,})\,$\ are the
initial and final wavefunctions for the intrinsic nuclear motion, respectively. The c.m. of relative motion between the nuclei is described by the coordinate $\bf R$ and  $\bf r$ is the intrinsic coordinate of the wavefunction for the nucleus of interest (usually the projectile in heavy ion charge exchange with radioactive beams).

\begin{figure}[h]
\begin{minipage}{22pc}
\includegraphics[width=18pc]{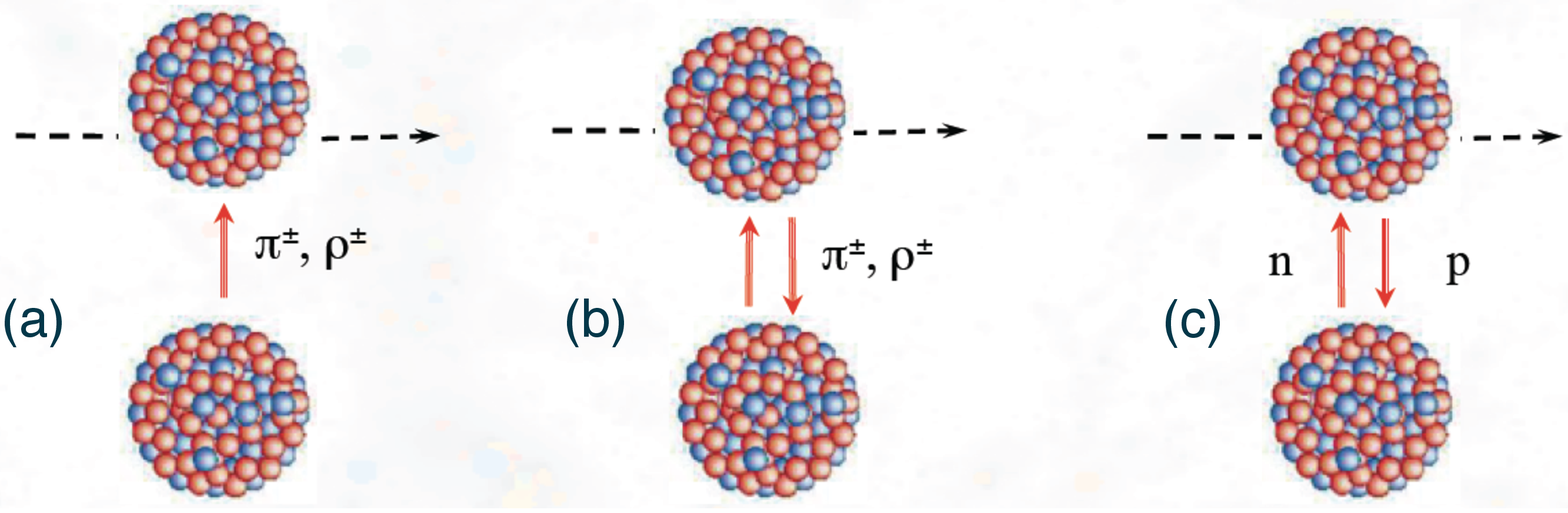}
\caption{\label{fig1}(a) One step charge exchange by means of $\pi^\pm$ and $\rho^\pm$ exchange. (b)  Two step charge exchange by means of $\pi^\pm$ and $\rho^\pm$ exchange. (c) Two step charge exchange by means of neutron and proton exchange.}
\end{minipage}\hspace{2pc}%
\begin{minipage}{14pc}
\includegraphics[width=14pc]{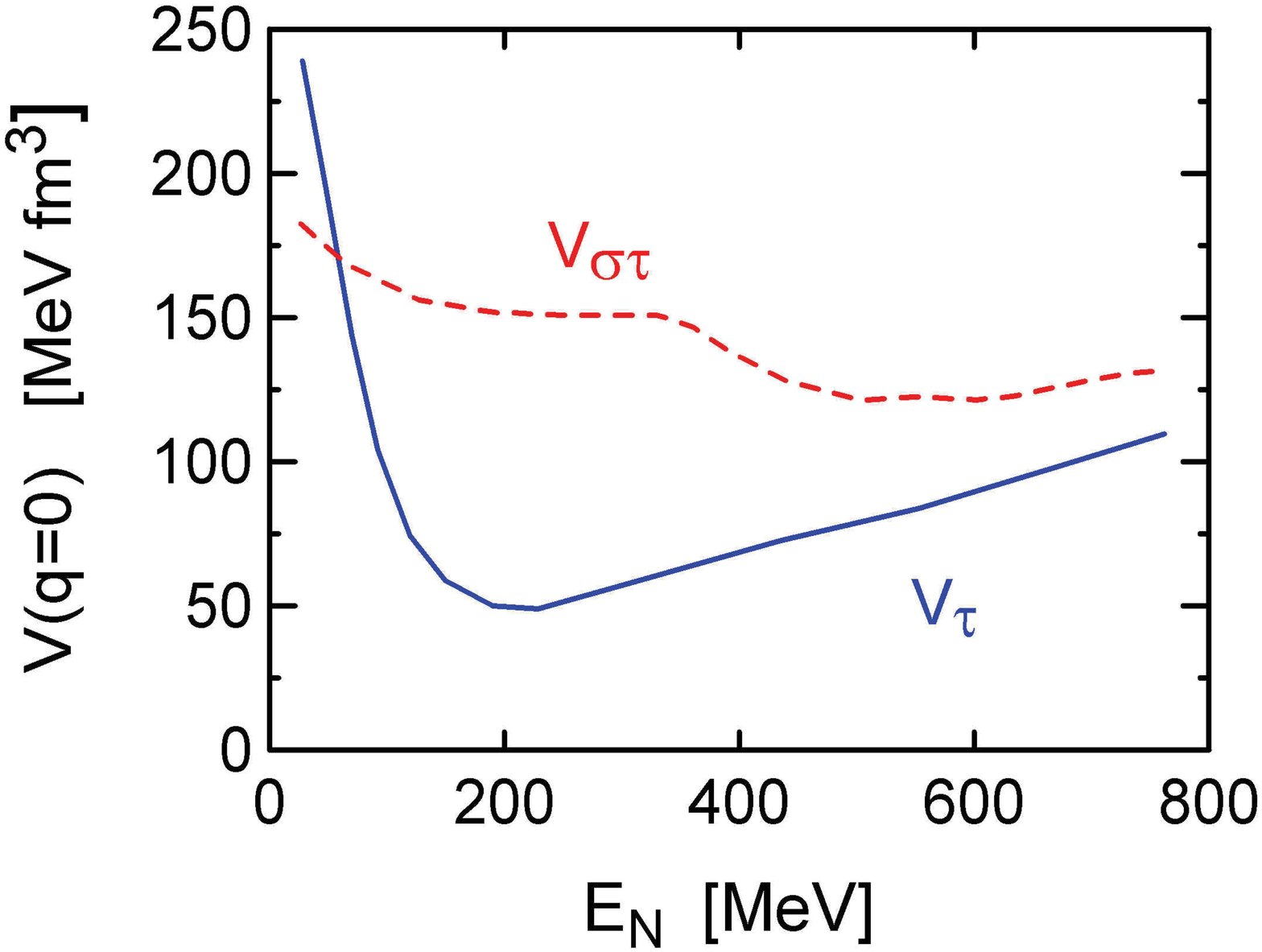}
\caption{\label{label}Spin and spin-isospin dependence of the effective nucleon-nucleon interaction used in direct reactions \cite{FL85}.}
\end{minipage} 
\end{figure}

In Eq. \ref{tfi1}, $\;V(\mathbf{R,r}\,)$  is the interaction potential
between the two nuclei which leads to the inelastic process $aA\rightarrow
bB$. It is taken as the interaction potential between a nucleon in nucleus $t$
($A$ or $B$) at position $\,\mathbf{r}_{t}\,$, with a nucleon in nucleus $p$
($a$ or $b$) at position $\,\mathbf{r}_{p}$. The meson exchange between these
nucleons is responsible for the charge exchange process in nucleus-nucleus
collisions at high energies ($E_{lab}>
50$ MeV/nucleon), for which the
physical exchange of nucleons is negligible due to severe phase-space limitations
\cite{Ber93}. 

We will now describe a method appropriate for inelastic scattering in high
energy collisions, along the lines of Ref. \cite{Ost92,Ber93,Dmit01}.  Even without details of the calculations, it is very easy to follow.
Using eikonal wavefunctions, the DWBA amplitude becomes
\begin{equation}
f_{aA\rightarrow bB}\left(  \theta\right)  =-\frac{m_{0}}{2\pi\hbar^{2}}%
\sum_{p,t}\int d^{3}R\left\langle \phi^{(b)}(\mathbf{r}_{p})\phi
^{(B)}(\mathbf{r}_{t})\left\vert V_{pt}(\mathbf{r})\,S\left(  b\right)
\exp\left[  -i\mathbf{q}\cdot\mathbf{R}\right]  \right\vert \phi
^{(a)}(\mathbf{r}_{p})\phi^{(A)}(\mathbf{r}_{t})\right\rangle, \label{tfi2}%
\end{equation}
where the sum includes all pairs of nucleons, with one from  nucleus $p$ and
another from nucleus $t$. The relative coordinates of the pair are related by
$
\mathbf{r}=\mathbf{R}+\mathbf{r}_{p}-\mathbf{r}_{t}
$,
$\,\mathbf{q}=\mathbf{k\,}^{\prime}\mathbf{-k}\;$is the momentum
transfer, $\mathbf{k}$ and $\mathbf{k}^{\prime}$ are the initial and final
momentum in the center of mass, and $b=|\mathbf{R\times\hat{k}}|$ is often
interpreted as the impact parameter.

In Ref. \cite{Ber93} the charge exchange cross sections were studied in terms
of $\pi+\rho$ exchange interactions based on the OPEP potential. In-medium
corrections of the nucleon-nucleon interaction were not taken into account.
These effects have been extensively studied by Franey and Love \cite{FL85} and other authors.
The important parts of the interaction for the charge-exchange reaction are
the central and the tensor components. In general, the spin-orbit part of the
interaction produces a negligible contribution to the cross sections.
The (in-medium) nucleon-nucleon interaction may be written as%
\begin{eqnarray}
V_{pt}(\mathbf{r})  &  =V^{C}\left(  r\right)  +V_{\sigma}^{C}\left(
r\right)  \left(  \mbox{\boldmath$\sigma$}_{p}\cdot
\mbox{\boldmath$\sigma$}_{t}\right)  +\left[  V_{\tau}^{C}\left(  r\right)
+V_{\sigma\tau}^{C}\left(  r\right)  \left(  \mbox{\boldmath$\sigma$}_{p}%
\cdot\mbox{\boldmath$\sigma$}_{t}\right)  \right]  \left(
\mbox{\boldmath$\tau$}_{t}\cdot\mbox{\boldmath$\tau$}_{p}\right)
\nonumber\\
&  +\left[  V^{T}\left(  r\right)  +V_{\tau}^{T}\left(  r\right)  \left(
\mbox{\boldmath$\tau$}_{t}\cdot\mbox{\boldmath$\tau$}_{p}\right)  \right]
S_{pt}\left(  \widehat{\mathbf{r}}\right)  +V^{LS}\left(  r\right)
\mathbf{l\cdot}\left(  \mbox{\boldmath$\sigma$}_{t}%
+\mbox{\boldmath$\sigma$}_{p}\right)  , \label{LF1}%
\end{eqnarray}
where the tensor operator  is
given by $
S_{pt}\left(  \widehat{\mathbf{r}}\right)  =3\left(
\mbox{\boldmath$\sigma$}_{p}\cdot\widehat{\mathbf{r}}\right)  \left(
\mbox{\boldmath$\sigma$}_{t}\cdot\widehat{\mathbf{r}}\right)  -\left(
\mbox{\boldmath$\sigma$}_{p}\cdot\mbox{\boldmath$\sigma$}_{t}\right),
$ and the spin-orbit term contains the angular momentum operator 
$
\mathbf{l=}\left(  \mathbf{r}_{t}-\mathbf{r}_{p}\right)
\times\left(  \mathbf{p}_{t}-\mathbf{p}_{p}\right) /\hbar$,
often yielding a negligible contribution to the cross sections.
The effective nucleon-nucleon potential has also to include an
exchange operator to fully account for anti-symmetrization. Thus, the
interaction (\ref{LF1}) should be understood as
$
V_{pt}(\mathbf{r})\rightarrow V_{pt}(\mathbf{r})\left[  1+\left(  -1\right)
^{l}P^{x}\right], \label{LF2}
$
where $P^{x}$\ is the space exchange operator which changes $\mathbf{r}%
\rightarrow-\mathbf{r}$ on the right and $\left(  -\right)  ^{l}$\ (with $l=$
relative angular momentum in the nucleon-nucleon system) ensures anti-symmetrization.

Using a spherical basis ($\mu=0,\pm1$), the scalar products can be written as
\begin{eqnarray}
\mbox{\boldmath$\tau$}_{t}\cdot\mbox{\boldmath$\tau$}_{p}  &  =-\sum_{\mu
}\sqrt{3}\left\langle 1\mu1-\mu|00\right\rangle \tau_{t}^{\left(  \mu\right)
}\tau_{p}^{\left(  -\mu\right)  }=\sum_{\mu}\left(  -1\right)  ^{\mu}\tau
_{t}^{\left(  \mu\right)  }\tau_{p}^{\left(  -\mu\right)  }\label{ssop}\\
\mbox{\boldmath$\sigma$}_{t}\cdot\mbox{\boldmath$\sigma$}_{p}  &  =-\sum_{\mu
}\sqrt{3}\left\langle 1\mu1-\mu|00\right\rangle \sigma_{t}^{\left(
\mu\right)  }\sigma_{p}^{\left(  -\mu\right)  }=\sum_{\mu}\left(  -1\right)
^{\mu}\sigma_{t}^{\left(  \mu\right)  }\sigma_{p}^{\left(  -\mu\right)  },
\label{ttop}%
\end{eqnarray}
where%
\[
\tau^{\left(  \pm1\right)  }=\mp\frac{1}{\sqrt{2}}\left(  \tau_{x}\pm
i\tau_{y}\right)  ,\ \ \ \ \ \ \ \tau^{\left(  0\right)  }=\tau_{z}\ .
\]
The $\mbox{\boldmath$\tau$}_{t}\cdot\mbox{\boldmath$\tau$}_{p}$ operator is
responsible for isospin exchange via the combination of $\tau^{\pm}$ operators
in Eq. (\ref{ssop}). Likewise, the $\mbox{\boldmath$\sigma$}_{t}%
\cdot\mbox{\boldmath$\sigma$}_{p}$ is responsible for spin-flip interactions. Further, introducing
$
V^{C}\left(  r\right)     \equiv V_{00}^{0}\left(  r\right)
, \  V_{\tau}^{C}\left(  r\right)  \equiv V_{01}^{0}\left(  r\right)
,\  V_{\sigma}^{C}\left(  r\right)  \equiv V_{10}^{0}\left(  r\right)
\ 
V_{\sigma\tau}^{C}\left(  r\right)     \equiv V_{11}^{0}\left(  r\right)
,\ V^{T}\left(  r\right)  \equiv V_{10}^{2}\left(  r\right)
,\  V_{\tau}^{T}\left(  r\right)  \equiv V_{11}^{2}\left(  r\right)
$,
Eq. (\ref{LF1}) can be written in the following more compact form
\cite{Cook84}%
\begin{equation}
V_{pt}(\mathbf{r})=\sum_{({K=0,2)ST}}V_{ST}^{K}\left(  r\right)
C_{S}^{K}\ Y_{K}\left(  \widehat{\mathbf{r}}\right)  \cdot\left[
\mbox{\boldmath$\sigma$}_{t}^{S}\otimes\mbox{\boldmath$\sigma$}_{p}%
^{S}\right]  ^{K}\left[  \mbox{\boldmath$\tau$}_{t}^{T}\cdot
\mbox{\boldmath$\tau$}_{p}^{T}\right]  \ , \label{vfund}%
\end{equation}
where $K=0$ corresponds to the central force, $K=2$ to the tensor force, and
$S,T$ to the spin and isospin labels of the force. The constants $C_{S}^{K}%
$\ have values $C_{0}^{0}=\left(  4\pi\right)  ^{1/2}$, $C_{1}^{0}=-\left(
12\pi\right)  ^{1/2}$, $C_{0}^{2}=0$, and $C_{1}^{2}=\left(  24\pi/5\right)
^{1/2}$. When $S=0$, $\mbox{\boldmath$\sigma$}^{S}$\ is the unit operator, and
when $S=1$, it becomes the Pauli spin operator $\mbox{\boldmath$\sigma$}$.
Likewise for the isospin operator $\mbox{\boldmath$\tau$}^{T}$. 
In Eq. \ref{vfund} the following notation has been used
\begin{equation}
Y_{K}\left(  \widehat{\mathbf{r}}\right)  \cdot\left[
\mbox{\boldmath$\sigma$}_{t}^{S}\otimes\mbox{\boldmath$\sigma$}_{p}%
^{S}\right]  ^{K}=\sum_{\mu}\left(  -1\right)  ^{\mu}Y_{K,-\mu}\left(
\widehat{\mathbf{r}}\right)  \left[  \mbox{\boldmath$\sigma$}_{t}^{S}%
\otimes\mbox{\boldmath$\sigma$}_{p}^{S}\right]  ^{K\mu}\ ,
\end{equation}
and%
\begin{equation}
\left[  \mbox{\boldmath$\sigma$}_{t}^{S}\otimes\mbox{\boldmath$\sigma$}_{p}%
^{S}\right]  ^{K\mu}=\sum_{m}\left\langle SmS\left(  \mu-m\right)
|K\mu\right\rangle \sigma_{t}^{S\left(  m\right)  }\sigma_{p}^{S\left(
\mu-m\right)  }\ .
\end{equation}

Now, defining
\begin{eqnarray}
\widetilde{V}_{ST}^{K}\left(  p\right)  =4\pi\int V_{ST}^{K}\left(  r\right)
\ j_{K}\left(  pr\right)  \ r^{2}\ dr\ , \label{vptf3}%
\end{eqnarray}
it is straightforward to show that 
\begin{eqnarray}  f_{aA\rightarrow bB}\left(  \theta\right)  &=&-\frac{m_{0}}{(2\pi)^{4}%
\hbar^{2}}\int d^{3}p\,d^{3}R\,S\left(  b\right)  \ \exp\left[  -i(\mathbf{q}+{\bf p})%
\cdot\mathbf{R}\right]  \ \ \nonumber\\
&  \times&\left\langle \phi^{(b)}(\mathbf{r}_{p})\phi^{(B)}(\mathbf{r}%
_{t})\left\vert \exp\left[  -i\mathbf{p}\cdot(\mathbf{r}_{p}-{\bf r_t})\right] 
\widetilde{V}_{pt}(\mathbf{p})  \right\vert \phi^{(a)}(\mathbf{r}_{p})\phi^{(A)}(\mathbf{r}%
_{t})\right\rangle \nonumber\\
&  =&-\frac{m_{0}}{(2\pi)^{4}\hbar^{2}}\sum_{{(K=0,2)ST}}i^{K}%
C_{S}^{K}\ \int d^{3}p\,d^{3}R\,S\left(  b\right)  \ \exp\left[  -i(\mathbf{q}+{\bf p})%
\cdot\mathbf{R}\right] \widetilde
{V}_{ST}^{K}\left(  p\right)  Y_{K}\left(  \widehat{\mathbf{p}}\right)
\nonumber\\
&  \times&\left\langle \phi^{(b)}(\mathbf{r}_{p})\phi^{(B)}(\mathbf{r}%
_{t})\left\vert \exp\left[  -i\mathbf{p}\cdot(\mathbf{r}_{p}-{\bf r_t})\right]  \left[
\mbox{\boldmath$\sigma$}_{t}^{S}\otimes\mbox{\boldmath$\sigma$}_{p}%
^{S}\right]  ^{K}\left[  \mbox{\boldmath$\tau$}_{t}^{T}\cdot
\mbox{\boldmath$\tau$}_{p}^{T}\right]    \right\vert \phi^{(a)}(\mathbf{r}_{p})\phi
^{(A)}(\mathbf{r}_{t})\right\rangle . \nonumber \\ \label{ftheta3}%
\end{eqnarray}
This result is simple and is also {\it exact}, depending on the  following assumptions: (a) the validity of DWBA, (b) the use of eikonal wavefunctions and (c) an effective nucleon-nucleon interaction obeying all symmetries allowed. Let us them under scrutiny one by one:
\begin{itemize}
\item[a - ] {\it Validity of DWBA}. This is not questionable, as the reaction is a genuine perturbative process. In fact, DWBA is used in all experimental analysis of charge-exchange reactions.
\item[b - ]  {\it Validity of eikonal waves.} This also leaves for little room for discussion. Eikonal waves are known to describe accurately the distorted waves in high energy collisions involving low energy transfers and forward angles \cite{BD04}.    
\item[c- ] {\it Effective nucleon-nucleon interactions.} This is not a problem either.  Effective interactions are the basis of all approaches to nucleus-nucleus collisions. 
Earlier calculations used a single Yukawa
interaction with a range of 1 fm \cite{WD75}. A review of existing data
[(p,p') and (p,n) reactions] gave $V_{11}^{0}=12\pm2.5$ MeV \cite{Aus72} for
such a force. Later, Petrovich \textit{et al}. \cite{Gol76,PS77} proposed to
include a pseudopotential, i.e., a $\delta\left(  \mathbf{r}\right)  $\ force, to it. Also, effective interactions derived by fitting matrix
elements of a sum of Yukawas to G-matrix elements of phenomenological
nucleon-nucleon potentials were introduced \cite{Be77,HKT85,ATB83}. These
interactions involve more than one Yukawa potential in an effort to consider
different meson exchanges. A popular interaction, developed by Love and Franey \cite{FL85} uses the antisymmetrized matrix elements of
an effective potential to fit the nucleon-nucleon t-matrix in heavy ion collisions at high energies. The right-hand side of Figure \ref{fig1} shows the isospin and the spin-isospin dependence of the  interaction separately as a function of the bombarding energy. One sees that in the region around 200 MeV/nucleon  the spin-isospin part clearly dominates over the isospin one. In fact, most of charge-exchange experiments are carried out at about this energy. The high energy also validates the use of the eikonal distorted waves.
\end{itemize}

The differential cross section for charge exchange is obtained by an average
of initial spins and sum over final spins, i.e.,
\begin{equation}
\frac{d\sigma}{d\Omega}=\,\frac{1}{(2J_{a}+1)(2J_{A}+1)}\,\sum_{\mathrm{spins}%
}\left\vert f_{aA\rightarrow bB}\left(  \theta\right)  \right\vert ^{2}\ .
\end{equation}
Obviously, at forward angles  $q =2k\sin(\theta/2)\simeq 0$ and the integral in Eq. \ref{ftheta3} becomes even simpler. But the replacement of $q=0$ in Eq. \ref{ftheta3} only yields Eq. \ref{tadeucci} with additional simplifications. A closer inspection of Eq. \ref{ftheta3}  shows that, neglecting the $p$-dependence of the nucleon-nucleon interaction, $V_{ST}^{K}\left(  p\right)$, and of the angular momentum dependence in $Y_{K}\left(  \widehat{\mathbf{p}}\right)$, will lead to the result that ${\bf R} \simeq {\bf r}_p - {\bf r}_t$. Due to the strong absorption at small values of $b$ (small $R$), the reaction is peripheral and only rather large values of $R$ will contribute in the integral. This means that only the tail of the nuclear wavefunctions will be probed in heavy ion charge exchange reactions, as in any other direct reaction. Eq. \ref{tadeucci} is thus justified if we can neglect the momentum dependence in the prodcut ${V}_{ST}^{K}\left(  p\right)  Y_{K}\left(  \widehat{\mathbf{p}}\right)$. A much stronger approximation assumes the absence of refraction and absorption, such that $S\left(  \mathbf{b}\right)  \simeq1$. In this case the integral over $\mathbf{R}$ yields $\mathbf{q\simeq-p}$, and for $q \simeq 0$, we again recover Eq. \ref{tadeucci}.

A detailed study of the theoretical validity of Eq. \ref{tadeucci} was carried out in Ref. \cite{BL97}. It was found that the extraction of Gamow-Teller matrix elements from heavy ion charge exchange experiments can be misleading if a bettter treatment of the reaction mechanism is not taken into account. At forward angles a clear difference between exact calculations based on Eq. \ref{ftheta3} and Eq. \ref{tadeucci} have been found \cite{BL97}. This conclusion might seem surprising, because weak interaction strengths have routinely been extracted from (p,n) and heavy ion exchange reactions using the simple formulation based on the empirical Eq. \ref{tadeucci}.  

\section{Double-beta decay studies with heavy ion charge exchange reactions}
Double beta decay  in nuclei can occur in nuclei without emission of a neutrino, if the neutrino is a Majorana particle (i.e.,  its own antiparticle). Theorists believe that if the neutrino is a Majorana particle then it is easier to explain the origins of its small, but non-zero, mass. If neutrinoless double beta decay is observed, it will also provide the neutrino mass, since  its decay  rate is proportional to the square of the neutrino mass. Ordinary double beta decay with neutrino emission has been observed (see Ref. \cite{Ber12} and references therein), but up to now no experiment has reported positive results on neutrinoless beta-decay.  
Based on the V-A theory of weak interactions it is possible to show that the half-life of neutrinoless double beta decay can exceed by large that of ordinary double beta decay. For all such reasons, there is a large ongoing effort in the nuclear theory community to calculate reliably, or measure directly or indirectly, matrix elements relevant for double beta-decay. As with the case of electron capture matrix elements, spin and spin-isospin nuclear response needs to be assessed. 

Charge exchange reactions can play a vital role in extracting matrix elements for double beta-decay by means of the two-step process displayed  in part (b) on the left-hand side of Figure \ref{fig1}. The physical exchange of two nucleons is ruled out because the cross section is simply too small at the relevant heavy ion energies. Besides, the theoretical formalism for it is much more  involved than that for pion and rho exchange. In Ref. \cite{Ber93} it was estimated that double charge exchange total cross sections in heavy ion collisions are smaller by a factor $10^{-4} - 10^{-5}$  compared to single charge-exchange. If the peak of the differential cross section at forward angles for single charge exchange is of the order of millibarns, it was estimated that the equivalent value for double charge exchange is of the order of microbarns. This requires intense beams and excellent detection efficiency. With recent advances on both fronts, it is expected that a new line of research will focus on these reactions in the near future.

\section{Conclusions}

Indirect techniques for nuclear astrophysics using new radioactive beam facilities worldwide consist indisputably of one of the most interesting work in present nuclear physics research. In this short review, I have focused on charge exchange reactions and the extraction of Gamow-Teller matrix elements of relevance for electron capture in stars and for neutrino induced reactions. The purpose is to illustrate the level of understanding in nuclear reaction theory for this particular case. We know that empirical formulas are very useful in physics. And it seems that the empirical Eq. \ref{tadeucci} works reasonably well for charge exchange reactions. But is lacks  a solid physics basis in some situations. This leaves us in an uncomfortable position because physics is driven by the desire to understand physical phenomena from first principles.  

\section{Acknowledgments}
This work was partially supported by the  U.S. NSF Grant No. 1415656 and the U.S. DOE Grant No. DE-FG02-08ER41533.


\section*{References}
\begin{thebibliography}{9}
\bibitem{Heg03} Heger A, Fryer C L., Woosley S E, Langer N and Hartmann D H 2003 {\it Astrophys. J.} {\bf 591} 288
\bibitem{Hix03} Hix W R et al. 2003 {\it Phys. Rev. Lett.} {\bf 91} 201102
\bibitem{Pin00} Martínez-Pinedo G,  Langanke and K, Dean D 2000 {\it Astrop. J. Sup.} {\bf 126} 493
\bibitem{Goo80} Goodman C D, Austin Sam M, Bloom S D, Rapaport J and Satchler G R 1980 {\it The (p,n) Reaction and the Nucleon-Nucleon Force}  (Plenum, New York)
\bibitem{Pet80} Petrovich F 1980 ibid. 
\bibitem{Tad87} Taddeucci T N et al. 1987 {\it Nucl. Phys. } {\bf A469} 125
\bibitem{Ost92} Osterfeld F 1992  {\it Rev. Mod. Phys.} {\bf 64} 491
\bibitem{Ber93} Bertulani C A 1993 {\it Nucl. Phys.} {\bf A554} 493
\bibitem{BL97} Bertulani C A and Lotti P 1997 {\it Phys. Lett.} {\bf B402} 237
\bibitem{Lens89} Lenske H, Wolter H H, Bohlen H G 1989 {\it Phys. Rev. Lett.} {\bf 62}  1457
\bibitem{BD04}Bertulani C A and Danielewicz D 2004 {\it Introduction to Nuclear Reactions} (IOP Publishing, London)
\bibitem {FL85}Franey M A and  Love W G 1985 {\it Phys. Rev.} \textbf{C31}, 481
\bibitem {Dmit01} Dmitrev V F et al. 2001 {\it Phys. Rev.} \textbf{C}\textbf{65} 015803
\bibitem {Cook84}Cook J et al. 1984 {\it Phys. Rev.} \textbf{C}\textbf{30} 1538
\bibitem {WD75}Wharton W R and Debevec P T 1975 {\it Phys. Rev.} \textbf{C11} 1963
\bibitem {Aus72}  Austin Sam M and  Crawley G M 1972 {\it The Two-Body Force on Nuclei}  (Plenum, New York)
\bibitem {Gol76}Golin M,  Petrovich F and  Robson D,  1976 {\it Phys. Lett.} \textbf{B64} 253
\bibitem {PS77}Petrovich F and Stanley D 1977 {\it Nucl. Phys.}  \textbf{ A275} 487
\bibitem {Be77}Bertsch G, Borysowich J, McManus H, and  Love W G 1977 {\it Nucl. Phys.} \textbf{A284} 399
\bibitem {HKT85} Hosaka A,  Kubo K I, and Toki H 1986 {\it Nucl. Phys.} \textbf{A444} 76
\bibitem {ATB83}Anantaraman N, Toki H, and  Bertsch G F 1983 {\it Nucl. Phys.} \textbf{A398} 269
\bibitem{Ber12} Beringer J et al. 2012 {\it Phys. Rev. } {\bf D86} 010001
\end{thebibliography}
\end{document}